\documentclass[english,pre,onecolumn,amsmath,amssymb,superscriptaddress]{revtex4}
\usepackage{graphicx,color,soul}

\usepackage{graphicx}
\usepackage{comment}
\usepackage{amsmath,amssymb,amsthm} 
\usepackage{verbatim}
\usepackage{float}

\definecolor{strawberry}{rgb}{1.0,0.0,0.5}

\begin{document}

\title{Isomorphs in nanoconfined liquids}

\author{Benjamin M. G. D. Carter}
\affiliation{H.H. Wills Physics Laboratory, Tyndall Avenue, Bristol, BS8 1TL, UK.}
\affiliation{Bristol Centre for Functional Nanomaterials, Tyndall Avenue, Bristol, BS8 1TL, UK.}

\author{C. Patrick Royall}
\affiliation{H.H. Wills Physics Laboratory, Tyndall Avenue, Bristol, BS8 1TL, UK.}
\affiliation{School of Chemistry, University of Bristol, Cantock's Close, Bristol, BS8 1TS, UK}
\affiliation{Centre for Nanoscience and Quantum Information, Tyndall Avenue, Bristol, BS8 1FD, UK}

\author{Jeppe C. Dyre}
\affiliation{Department of Science and Environment, Roskilde University, Postbox 260, DK-4000 Roskilde, Denmark}

\author{Trond S. Ingebrigtsen}
\affiliation{Department of Science and Environment, Roskilde University, Postbox 260, DK-4000 Roskilde, Denmark}

\begin{abstract}
We study in this paper the possible existence of Roskilde-simple liquids and their isomorphs in a rough-wall nanoconfinement. Isomorphs are curves in the thermodynamic phase diagram along which structure and dynamics are invariant in suitable nondimensionalized units. Two model liquids using molecular dynamics computer simulations are considered: the single-component Lennard-Jones (LJ) liquid and the Kob-Andersen binary LJ mixture, both of which in the bulk phases are known to have isomorphs. Nanoconfinement is implemented by adopting a slit-pore geometry with fcc crystalline walls; this implies inhomogenous density profiles both parallel and perpendicular to the confining walls. Despite this fact and  consistent with an earlier study [Ingebrigtsen \textit{et. al}, Phys. Rev. Lett. \textbf{111}, 235901 (2013)] we find that these nanoconfined liquids have isomorphs to a good approximation. More specifically, we show good scaling of inhomogenous density profiles, mean-square displacements, and higher-order structures probed using the topological cluster classification algorithm along the isomorphs. From this study, we conjecture that in experiments, Roskilde-simple liquids may exhibit isomorphs if confined in a suitable manner, for example with carbon nanotubes. Our study thus provides an alternative framework for understanding nanoconfined liquids.
\end{abstract}

\date{\today}

\maketitle

\section{Introduction}

An important simplification in the study of liquids via computer simulations is to apply so-called periodic boundary conditions. The liquid is thereby free from any confining surfaces which affect its 
structure and dynamics by imposing an external force field on the liquid \cite{tildesley, frenkel}. This simplification is, however, hard to achieve in experiments, and most liquids in nature are in contact with, or confined by, one or several surfaces \cite{tox1981,schoen1987,schoen1988,drake1990,granick1991,jackson1991,keddie1994,bhushan1995,scheidler2000,starr2002,alba2006,richert2011}.
Recent experiments on levitation of metallic alloys using electrostatic or magnetic fields \cite{voightmann2008,meyer2010,brillo2011,yang2014}, ionic solution droplets in optical tweezers \cite{robinson2020}, and especially colloids \cite{williams2015jcp} come closer to this simplification of standard computer simulations \cite{voightmann2008,meyer2010,brillo2011,yang2014}, but naturally still have ''free surfaces'' that may affect the probed quantities. Significantly, the added complexity induced by the walls
has made fundamental theories of nanoconfined liquids slower to develop, in particular for the dynamics of nanoconfined liquids \cite{evan1979,krakoviack2005,lang2010,lang2012,hansen2013,hansen2015}.

Roskilde-simple liquids (also called R-simple liquids) are liquids with strong correlations between equilibrium fluctuations of the virial $W$ and the potential energy $U$ in the NVT ensemble \cite{paper1,paper2,paper3,paper4,paper5,dyre2014,thomas2014,ingebrigtsen2015}. Van der Waals and metallic liquids have been shown to belong to this class of liquids whereas, e.g., hydrogen-bonding liquids are not R-simple. R-simple liquids have isomorphs in the thermodynamic phase diagram which are curves along which structure and dynamics are
invariant in reduced units. This fact makes R-simple liquids simpler than other types of liquids. As an example, 
Rosenfeld's excess entropy scaling can be explained using the concept of isomorphs \cite{rosenfeld1, rosenfeld2}. In Rosenfeld's excess-entropy scaling reduced transport coefficients
are functions of the entropy minus the ideal contribution at the same density and temperature, i.e., $\widetilde{X} = f(S_{\rm ex})$, where $S_{\rm ex}(\rho, T) = S(\rho, T) - S_{\rm id}(\rho, T)$.
Since both the reduced dynamics and the excess entropy are invariant along the same curves (isomorphs) this fact explains Rosenfeld's excess entropy scaling. This scaling law is, however, only one of many consequences of having isomorphs (see, e.g., Ref. \cite{paper4}).

Extending the isomorph theory to nanoconfined fluids is therefore of paramount importance as this would offer an alternative framework in  
which confined fluids could be understood and analyzed. An earlier computer simulation study investigated R-simple liquids in confinement using an idealized slit-pore geometry \cite{ingebrigtsen2013}. It was found that even heavily nanoconfined liquids have isomorphs to a good approximation, except when the confinement is around one or two particle diameters. Idealized slit-pore confinement implies that only the density profile perpendicular to the walls is inhomogenous.

To model more realistic confinement conditions we study in this paper two model liquids confined to a slit-pore geometry with fcc crystalline walls. The structure and dynamics of liquids confined by fcc walls have been studied before and shown to exhibit density profiles that are highly inhomogenous, both parallel and perpendicular to the confining walls \cite{tox1981,schoen1987,schoen1988}. Our aim here is to investigate whether isomorphs survive under such strong inhomogeneities. This is a first step in the direction of studying more realistic confinement conditions relevant, e.g., for industrial applications and biological systems.

We find that, despite the apperance of strong inhomogenous density profiles in the liquid, isomorphs do survive down to a few particle-diameters confinements, enabling the applicability
of results from the isomorph theory to more complex confined liquids. We conjecture from this study that R-simple liquids and isomorphs are relevant for a much larger class of confinements, consistent with studies of excess-entropy scaling in nanoconfinement \cite{truskdumbbellbulk}.

At lower temperatures than what we consider here, higher-order structures have been shown to exhibit differences between isomorphic states \cite{malins2013iso}. Here therefore, we also probe measures of higher-order structure and investigate minimum energy clusters in the liquid of interest \cite{malins2013tcc}. We find that little difference is seen in higher-order structure along isomorphs of the confined liquids.

The paper is organised as follows. Section \ref{MD} introduces the models and methods we apply in this study and Sec. \ref{sum} gives a short introduction to R-simple liquids and their isomorphs. Section \ref{res1} presents results for the single-component Lennard-Jones (LJ) liquid where we, amongst other things, study isomorphs. Section \ref{res2} presents similar results
for the Kob-Andersen binary LJ mixture. Section \ref{con} summaries and presents a brief outlook. 

\section{Simulation methods}\label{MD}

We use standard Nos\'e-Hoover molecular dynamics computer simulations in the NVT ensemble (the RUMD package \cite{rumd}) to study two model liquids in confinement: the single-component Lennard-Jones (SCLJ) liquid and the Kob-Andersen binary LJ mixture (KABLJ) \cite{ka1,ka2}. 
In both models, the pair interaction between the liquid particle $i$ of type $\alpha$ and the liquid particle $j$ of type $\beta$ is described by the LJ pair potential given by

\begin{equation}
v_{\alpha \beta}(r) = 4\epsilon_{\alpha \beta} \Big[\Big(\frac{\sigma_{\alpha \beta}}{r}\Big)^{12} - \Big(\frac{\sigma_{\alpha \beta}}{r}\Big)^6\Big]\label{lj},
\end{equation}
where $\epsilon_{\alpha \beta}$ is the strength of the pair interaction ($\alpha$ or $\beta$ is equal to type $A$ or $B$ for KABLJ), $r$ is the distance separating the particles, and $\sigma_{\alpha \beta}$ is the separation distance at which the pair potential is zero. For the LJ model we have $\epsilon_{\rm AA}$ = 1, $\sigma_{\rm AA}$ = 1, and $m_{\rm A}$ = 1, whereas the KABLJ mixture has $\sigma_{\rm AA}$ = 1, $\epsilon_{\rm AA}$ = 1, $\sigma_{\rm AB}$ = 0.80, $\epsilon_{\rm AB}$ = 1.5,
$\sigma_{\rm BB}$ = 0.88, and $\epsilon_{\rm BB}$ = 0.5. The masses of both particles in the KABLJ model are unity. The pair potential is truncated-and-shifted at the distance $r_{c}$ = 2.5$\sigma_{\alpha \beta}$.

The number of liquid particles is for SCLJ and KABLJ: $N$ = 2000$-$4000 and  $N$ = 2500$-$4800, respectively. For most state points we simulate around one million time steps with a time step of $\Delta t$ = 0.0025 after obtaining equilibrium. Equilibrium is ascertained from the decay of the intermediate scattering function, and running the simulations back-to-back at least twice.

\subsection{Simulation units}

Throughout the study, we use two different sets of nondimensionalized units: One set is based on the microscopic parameters of the LJ potential
with length scale $\sigma_{\rm AA}$ and energy scale $\epsilon_{\rm A A}$ of the larger (A) particle, which is standard in computer simulations, and another set
of nondimensionalized units using macroscopic quantities with length given in units $\rho^{-1/3}$, energy in units of $k_{\rm B}T$, and time in
units of $\rho^{-1/3}\sqrt{m/k_{\rm B}T}$ ($m$ is the particle mass) as applied in isomorph scaling \cite{paper4}. We refer to macroscopic
nondimensionalized units as reduced units and use a tilde above the variable name to indicate a reduced quantity; otherwise LJ units are implicitly assumed.

\subsection{Nanoconfinement}

\begin{figure}[H]
  \centering
  \includegraphics[width=80mm]{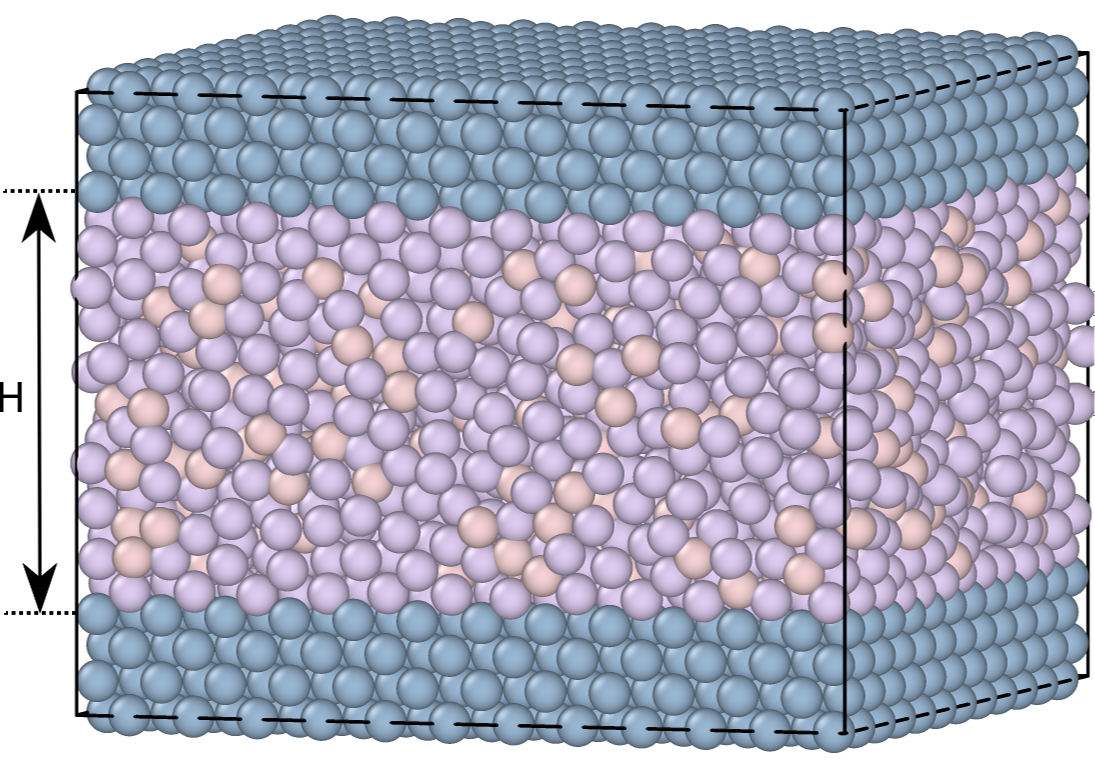}
  \caption{The simulated slit-pore geometry confinement for the KABLJ mixture \cite{ovito}. Dark purple is the larger A-particle,
    pink is the smaller B-particle, and blue is the crystalline wall particles. The distance between the two walls is $H$, here $H$ = 8.}\label{fig1}
\end{figure}

Nanoconfinement is modelled using a slit-pore geometry in which the 100 surface of an fcc crystal is exposed to the liquid. The two crystal planes are in registry (i.e., out of sync). The distance between the two walls is denoted by $H$, measured from the
centers of the confining fcc particles (see Fig. \ref{fig1}).

The liquid-wall pair interactions are also described by the LJ potential in Eq. (\ref{lj}). For the SCLJ liquid we use the parameters
$\sigma_{\rm AW}$ = 1 and $\epsilon_{\rm AW}$ = 1, in which $W$ denotes a wall particle. For the KABLJ mixture we derived the interaction parameters from Lorentz-Berthelot mixing rules with $\sigma_{\rm AW}$ = 0.97, $\epsilon_{\rm AW} = 1$,  $\sigma_{\rm BW}$ = 0.91, and $\epsilon_{\rm BW}$ = 0.707, where we used $\sigma_{\rm WW}$ = 0.94 and  $\epsilon_{\rm WW}$ = 1 to calculate these numbers. The cutoff of the liquid-wall pair interaction is $r_{c}$ = 2.50$\sigma_{\rm \alpha W}$. The density of the fcc walls is for the majority of the simulations kept fixed at $\rho_{\rm W}$ = 1, but we study also the effect of varying this parameter. The walls consist of around $N$ = 1600 particles.

For the SCLJ liquid we use the density $\rho$ = 0.85 as a reference with $T$ = [0.70, 10] and $H$ = [2, 10]. For KABLJ we focus on $\rho = 1.20$ with $T$ = [0.7, 12] and $H$ = [4, 10]. These densities are standard for studies of bulk liquids. 
The simulations use a Nos\'e-Hoover NVT thermostat on the liquid particles whereas, for simplicity, the wall particles are frozen in place, i.e., $T$ = 0 for the walls. Thermostatting the walls and the nature of the exposed surface are known to have observable effects on the structure and dynamics of confined liquids \cite{tox1981,schoen1987,schoen1988}. The phase behavior of this particular type of confinement has been studied in detail elsewhere see, e.g., Ref. \cite{dominguez2003}.

In sections \ref{res1} and \ref{res2}, we compare results of the topological cluster classification \cite{malins2013tcc} for the confined system with the bulk system at the same state points (density and temperature). For these simulations we use standard Monte-Carlo simulation with $N=4000$ particles for the models specified above.

\section{Roskilde-simple liquids}\label{sum}

We provide here a brief introduction to Roskile-simple (R-simple) liquids and their isomorphs; a review is given in Ref. \onlinecite{dyre2014}. R-simple liquids are characterized by
strong correlations between the equilibrium fluctuations of the potential energy $U$ and virial $W$ (recall $PV = N k_{\rm B}T + W$)  in the NVT ensemble \cite{paper1,paper2,paper3,paper4,paper5,dyre2014,thomas2014,ingebrigtsen2015}. This correlation is quantified by the Pearson correlation coefficient $R$ defined by \cite{paper1}

\begin{align}
  R & = \frac{\langle \Delta W \Delta U \rangle}{\sqrt{\langle (\Delta W)^2 \rangle \langle (\Delta U)^2 \rangle}}.\label{Rdef}   
\end{align}
Here $\Delta$ denotes deviation from the average value, and the averages are taken in the NVT ensemble (i.e., canonical ensemble averages). R-simple liquids are those for which the correlation coefficient $R$ is above 0.90, a criterion that depends on the state point. The correlation coefficient has been shown to be high in large parts of the phase diagram for many systems, typically in the condensed liquid and solid phases, but not in the gas phase. 

Van der Waals and metallic liquids are usually R-simple whereas, e.g., hydrogen-bonding, covalent-bonding and strongly ionic liquids are not.
More specifically, in simulations the SCLJ liquid, the KABLJ mixture, the Wahnstr{\"o}m OTP model, bead-spring polymer models, and many more all belong to this class of liquids. Strong $UW$ correlation
has also been verified in experiments on weakly dipolar organic molecules \cite{gammagamma,roed2013,wence2014}.

Initially, the bulk liquid phase was studied, but the concept of R-simple liquids was later extended to crystals \cite{crystals}, nanoconfined liquids \cite{ingebrigtsen2013}, nonlinear sheared liquids \cite{sllod}, polydisperse liquids \cite{ingebrigtsen2015,ingebrigtsen2016}, quantum-mechanical \textit{ab inito} liquid metals \cite{metals}, and more \cite{pedersen2016}. 

R-simple liquids are characterized by the following ordering of potential energy values \cite{thomas2014}

\begin{align}
 U(\textbf{R}_{\rm a}) < U(\textbf{R}_{\rm b}) \Rightarrow U(\lambda\textbf{R}_{\rm a}) < U(\lambda\textbf{R}_{\rm b}),\label{iso}
\end{align}
where $\textbf{R}_{\rm a}$ and $\textbf{R}_{\rm b}$ are 3N-dimensional configurational-space vectors of a given density and $\lambda$ is a 
factor scaling uniformly
these configurations to a new density. 

R-simple liquids exhibit a number of simple properties \cite{PRX}, most of which are consequences of the existence of isomorphs. Isomorphs are curves in the thermodynamic phase diagram of R-simple liquids along which structure and dynamics to a good approximation are invariant in reduced units (see previous section). Isomorphs are defined as curves of constant excess entropy and in simulations can be generated via the following relation which keeps the excess entropy constant

\begin{align}
  \gamma & \equiv \Big(\frac{\partial \ln T}{\partial \ln \rho}\Big)_{S_{\rm ex}} = \frac{\langle \Delta W \Delta U \rangle}{\langle (\Delta U)^2 \rangle}. \label{test1}
\end{align}
The equation is a general thermodynamic relation in the NVT ensemble \cite{paper4,sexreview}. The parameter $\gamma$ is called the density-scaling exponent because it is a key quantity when applying density scaling \cite{tolle, dreyfus2003, paper4}. The procedure to generate an isomorph in simulations using the above relation is as follows: A simulation is performed at a given state point, $\gamma$ is calculated, a new slightly higher or lower density is chosen, and from discretization of Eq. (\ref{test1}) the new temperature is calculated.

In this article we generate isomorphs using a different procedure. A first-order approximation to Eq. (\ref{iso}) implies that the Boltzmann factors of two isomorphic state points are proportional (also the old definition of isomorphs \cite{paper4}), i.e.,

\begin{align}
\exp\big(-U(\textbf{R}^{(1)})/k_{\rm B}T_{\rm 1}\big) = C_{12}\exp\big(-U(\textbf{R}^{(2)})/k_{\rm B}T_{\rm 2}\big),
\end{align}
where $C_{\rm 12}$ is constant and the comparison is performed for configurations for which $\rho_{1}^{1/3}\textbf{R}^{(1)}$ = $\rho_{2}^{1/3}\textbf{R}^{(2)}$, i.e., having the same reduced coordinates. From this equation it follows that if a simulation is performed at density $\rho_{\rm 1}$ and temperature $T_{\rm 1}$ and configurations are scaled uniformly to a different density
$\rho_{\rm 2}$ at which the potential energy is evaluated, the linear regression slope of $U_{2}$ vs $U_{1}$ provides the ratio of the temperatures $T_{\rm 2}/T_{\rm 1}$ of the two isomorphic state points. This is called the direct isomorph check \cite{paper4}. In this article we change density in steps of approximately 5\%. For confined systems the wall distance $H$ is an independent variable, similar to density and temperature in bulk liquids. We choose here to let $H$ follow the overall scaling in density, i.e. $H$ scales with $(\rho_{2}/\rho_{1})^{1/3}$. This choice is consistent with the definition in a previous study of isomorphs in nanoconfinement \cite{ingebrigtsen2013}. This study \cite{ingebrigtsen2013,trond} also indicated that it might be possible to keep the walls fixed but we do not consider this in more detail here. It is important to note that the state points we identify as being isomorphic under confinement in general are not isomorphic in the bulk.

\section{Single-component Lennard-Jones liquid}\label{res1}

We commence the study by probing the correlation coefficient $R$ and density-scaling exponent $\gamma$ for the SCLJ liquid in confinement. The next section considers the same quantities for the KABLJ mixture.

\subsection{Variation in the correlation coefficient $R$ and density-scaling exponent $\gamma$}

This section studies how the above mentioned quantities are affected by changing the following parameters related to the confinement: the distance between the two walls $H$, the strength of the liquid-wall interaction $\epsilon_{\rm LW}$, and the surface roughness $\rho_{\rm W}$.

\begin{figure}[H]
  \centering
  \includegraphics[width=140mm]{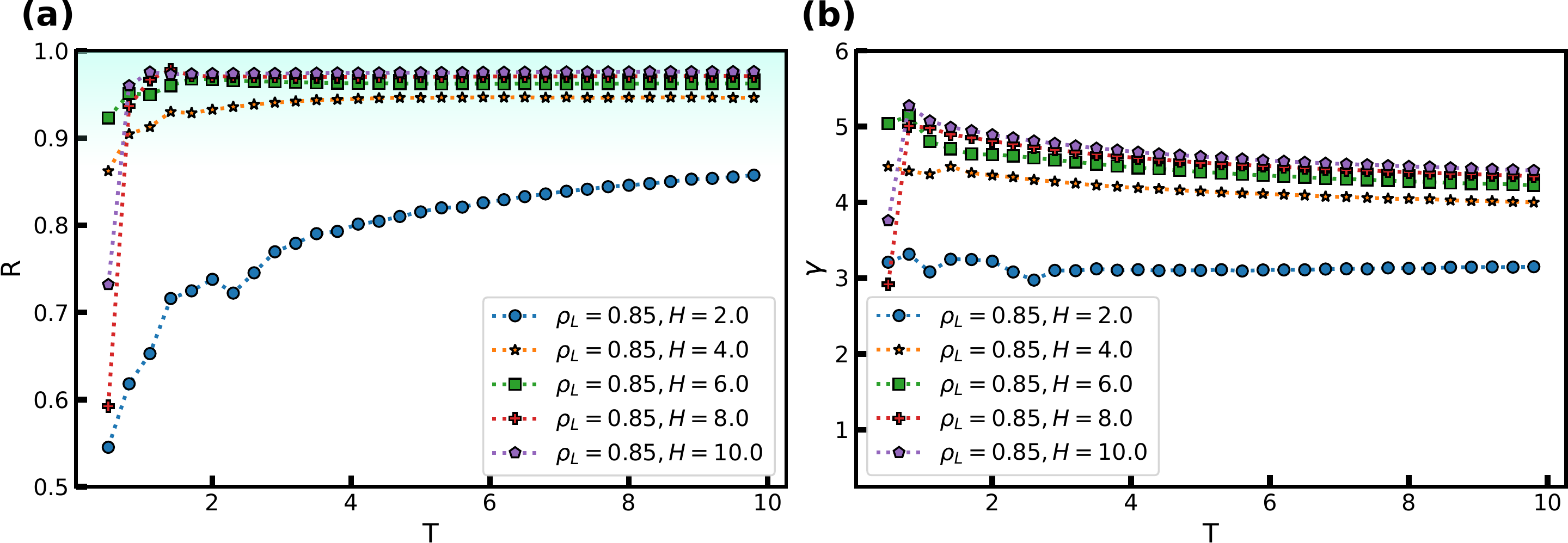}
  \caption{The correlation coefficient $R$ and density-scaling exponent $\gamma$ as a function of temperature $T$ for several slit-pore widths $H$ for the SCLJ liquid.
    The liquid density is $\rho_{\rm L} = 0.85$ and the wall density is $\rho_{\rm W} = 1$. (a) $R$ as a function of temperature. (b) $\gamma$ as a function of temperature.}\label{fig2}
\end{figure}

Figure \ref{fig2} shows how $R$ and $\gamma$ (Eqs. (\ref{Rdef}) and (\ref{test1})) vary with temperature for several slit-pore widths in the range $H$ = [2, 10].  The confinement thus ranges
from almost a single layer of liquid particles to more bulk-like conditions. The densities of the liquid and wall are kept constant with $\rho_{\rm L} = 0.85$ and $\rho_{\rm W} = 1$, respectively.
For simplicity, we here and henceforth define the liquid density as
$\rho_{\rm L} \equiv N / AH$, where $A$ is the exposed surface area of the crystal. We thus do not take into account any excluded volume near the walls when calculating the confined liquid
density \cite{truskdumbbellbulk}. As a reference, the bulk liquid has $R \approx 0.96$ at the chosen liquid density.

The wall separation $H$ = 2 shows markedly different behavior from the other slit-pore widths having $R < 0.90$ for all temperatures. This is consistent with an earlier study that also
observed breakdown of R-simple liquids for wall separations close to a single particle layer \cite{ingebrigtsen2013}. However, as the slit-pore width is increased $R$ also increases and already at $H$ = 4, which still corresponds to a strongly 
confined liquid, we find good correlation between the virial and potential energy, i.e., $R > 0.90$. Depending on $H$, we find an increase in $R$ with temperature as close encounters start to play a bigger role. We find also some irregularities in these observations for low temperatures due to crystallisation in the layer closest to the wall.

The density-scaling exponent $\gamma$ displays the opposite trend in Fig. \ref{fig2}(b) with a monotonic decrease with temperature for all $H$. $\gamma$ is noted to increase with slit-pore width, but no theory currently exists for how $\gamma$ should depend on $H$, as is the case for bulk liquids \cite{paper2, thermoscl}. We find a possible slit-pore-width dependent plateau for $\gamma$ between 4 and 5, which signifies that $\gamma$ is also dependent on $H$ in confined liquids. More investigations are nevertheless needed to determine if this $H$-dependent plateau truly exists and if it has a physical significance.

\begin{figure}[H]
  \centering
  \includegraphics[width=140mm]{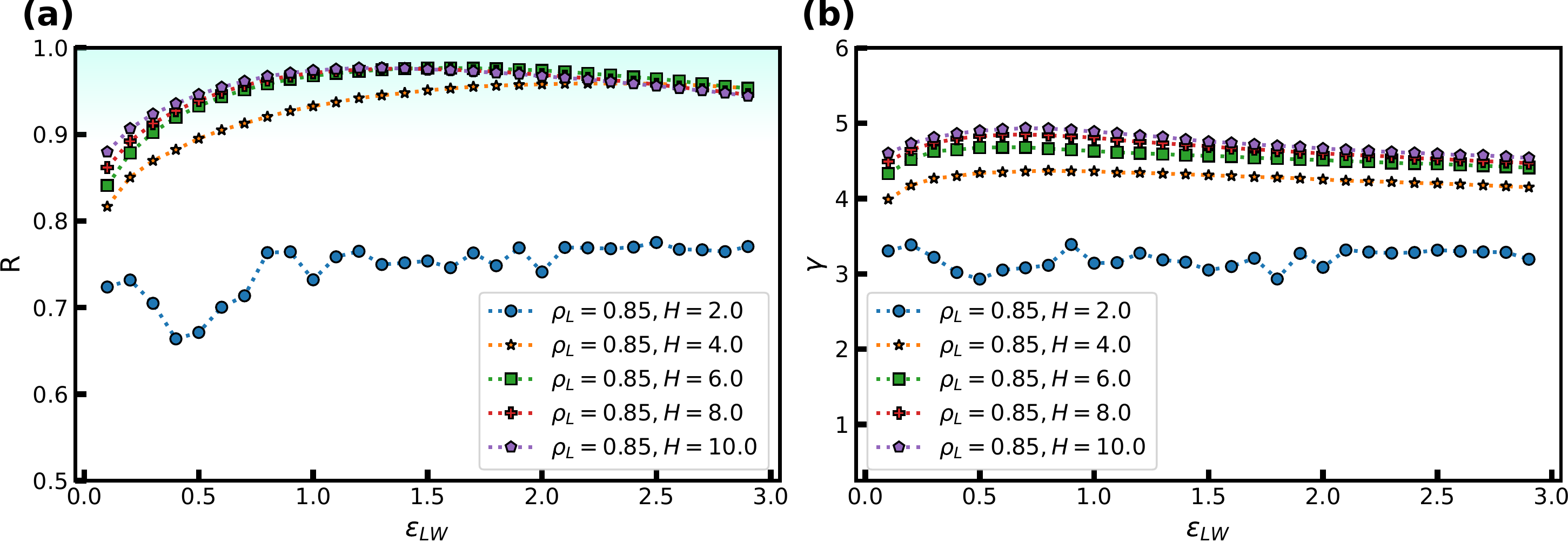}
  \caption{The correlation coefficient $R$ and density-scaling exponent $\gamma$  as a function of the strength of the liquid-wall interaction $\epsilon_{\rm LW}$ for several slit-pore widths $H$. Temperature is fixed with $T$ = 2,
    the liquid density is $\rho_{\rm L}$ = 0.85, and the wall density is $\rho_{\rm W}$ = 1. (a) R as a function of $\epsilon_{\rm LW}$. (b) $\gamma$ as a function of $\epsilon_{\rm LW}$. }\label{fig3}
\end{figure}
We now consider the effect of varying the attraction between the wall and the confined liquid particles.
Figure \ref{fig3} shows $R$ and $\gamma$ as a function of $\epsilon_{\rm LW}$ at $T$ = 2 and $\rho$ = 0.85, again for several $H$. As the simulations use $\epsilon_{\rm AA} = 1$, for $\epsilon_{\rm LW} > 1$ we have an ``attractive'' wall and when $\epsilon_{\rm LW} < 1$ it becomes a ``repulsive''' wall with respect to the interactions between the liquid particles.

Figure \ref{fig3}(a) shows that $R$ depends significantly on $\epsilon_{\rm LW}$. Depending on $H$, the correlation coefficient $R$ may both decrease and increase when
the wall becomes more attractive or repulsive. The interplay between the liquid-liquid and liquid-wall interactions is thus highly nontrivial.
The density-scaling exponent in Fig. \ref{fig3}(b) displays a behavior that mimics
that of the correlation coefficient, but with the maximum displaced to lower values of $\epsilon_{\rm LW}$. 

The highest value of $R$, and therefore the maximum, is expected to appear when the wall particles are most similar to the liquid particles, which means here around $\epsilon_{LW} = 1$. 
In spite of this, we observe that the maximum occurs somewhat to the right around $\epsilon_{LW} = 1.5$ for $R$ and
the opposite for $\gamma$. We currently have no explanation for why this is the case.

\begin{figure}[H]
  \centering
  \includegraphics[width=140mm]{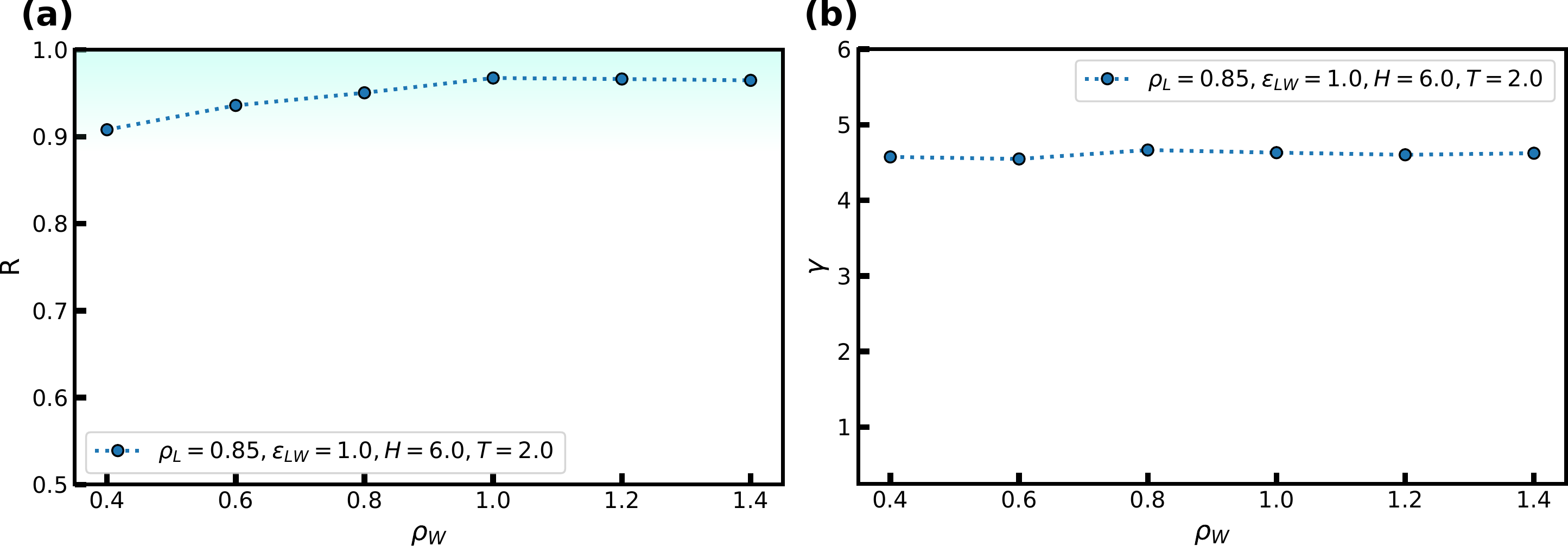}
  \caption{The correlation coefficient $R$ and density-scaling exponent $\gamma$ as a function of surface roughness varied by changing the wall density $\rho_{\rm W}$ at $T$ = 2, $H$ = 6, and $\epsilon_{\rm LW}$ = 1. The liquid density is $\rho_{\rm L}$ = 0.85. (a) $R$
    as a function of $\rho_{\rm W}$. (b) $\gamma$ as a function of $\rho_{\rm W}$.}\label{fig4}
\end{figure}

Another means to probe the coupling between the confined liquid and the walls is to change the density of the fcc walls $\rho_{\rm W}$.
In Fig. \ref{fig4} we show the correlation coefficient and density-scaling exponent  as a function of the ''surface roughness'' of the crystal (i.e., as a function of $\rho_{\rm W}$). The liquid density is kept constant at $\rho_{\rm L}$ = 0.85 and $H$ = 6.

We observe for low $\rho_{\rm W}$, i.e., when the surface roughness is high, that $R$ decreases when the density of the wall particles is reduced, though it remains above 0.90. This effect can be attributed to particles
penetrating into the walls (not shown). For high $\rho_{\rm W}$ the correlation cofficient remains virtually constant. Almost no  effect on $\gamma$ is noted in Fig. \ref{fig4}(b) for both
low and high $\rho_{\rm W}$.

\subsection{Isomorph invariance of reduced density profiles and dynamics}

We now turn our attention to  isomorphs in the nanoconfined system. To this end, we consider the behaviour along an isomorph and contrast it with an isochore at a density $\rho_{\rm L} = 0.85$ as
the isomorph concept is approximate. Isomorphs in the SCLJ liquid were generated by the direct-isomorph-check method (see Sec. \ref{sum}). We consider two different slit-pore widths, one with $H \approx$ 6
and one with $H \approx$ 10; recall that $H$ is adjusted with the liquid density along isomorphs. These two distances span a strong and medium confined liquid. 

\begin{figure}[H]
  \centering
  \includegraphics[width=140mm]{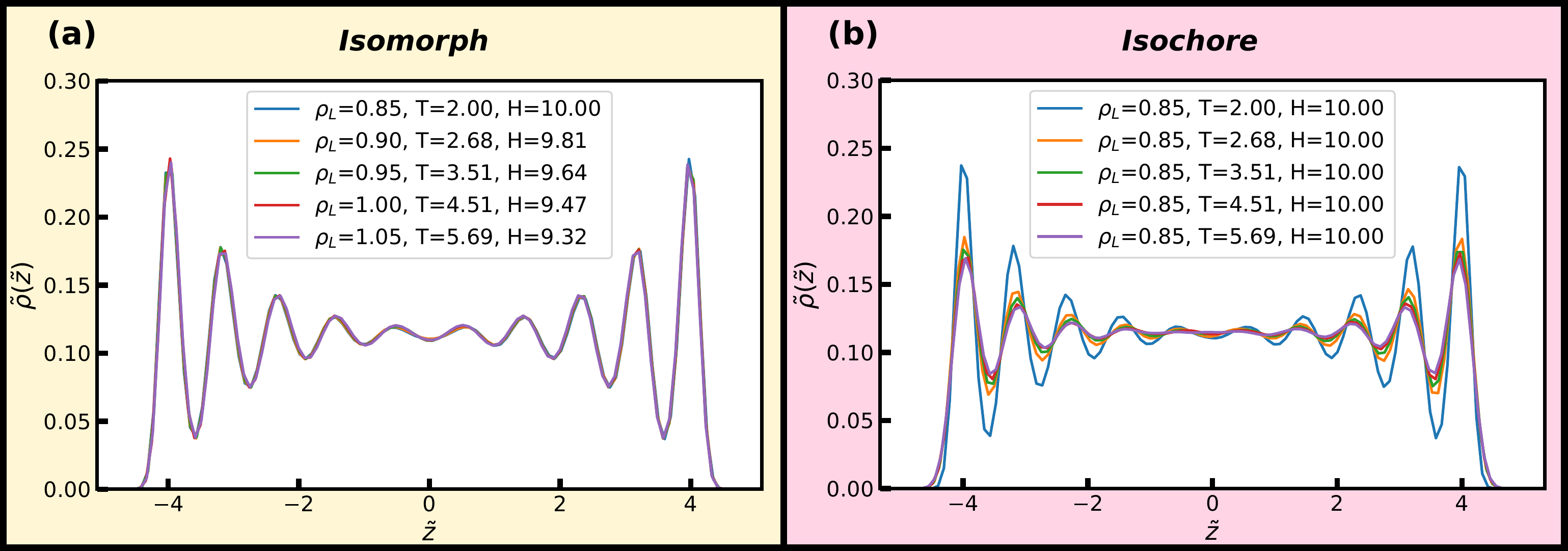}
  \caption{Reduced density profiles of the liquid particles perpendicular to the walls along an isomorph and an isochore.  Excellent invariance is seen along the isomorph but not along the isochore with significant changes in all peak heights. (a) Isomorph. (b) Isochore.}\label{fig5}
\end{figure}

Density profiles perpendicular to the walls along an isomorph with $H \approx$ 10 are shown in Fig. \ref{fig5}(a). For comparison, results for an isochore with the same temperature variation are given in Fig. \ref{fig5}(b). From this point forward a yellow figure background denotes data obtained along an isomorph and a pink figure background denotes data obtained along an isochore. 

We find that the density profiles to a good approximation are invariant along the isomorph with a 24\% density increase, whereas this is seen not to be the case along the isochore in Fig. \ref{fig5}(b). For the isochore, significant changes in all peak heights with temperature are observed, in particular for the layer closest to the wall indicating pre-crystallisation even at $H$ = 10.
This pre-crystallization is, however, nicely preserved on the isomorph.

\begin{figure}[H]
  \centering
  \includegraphics[width=120mm]{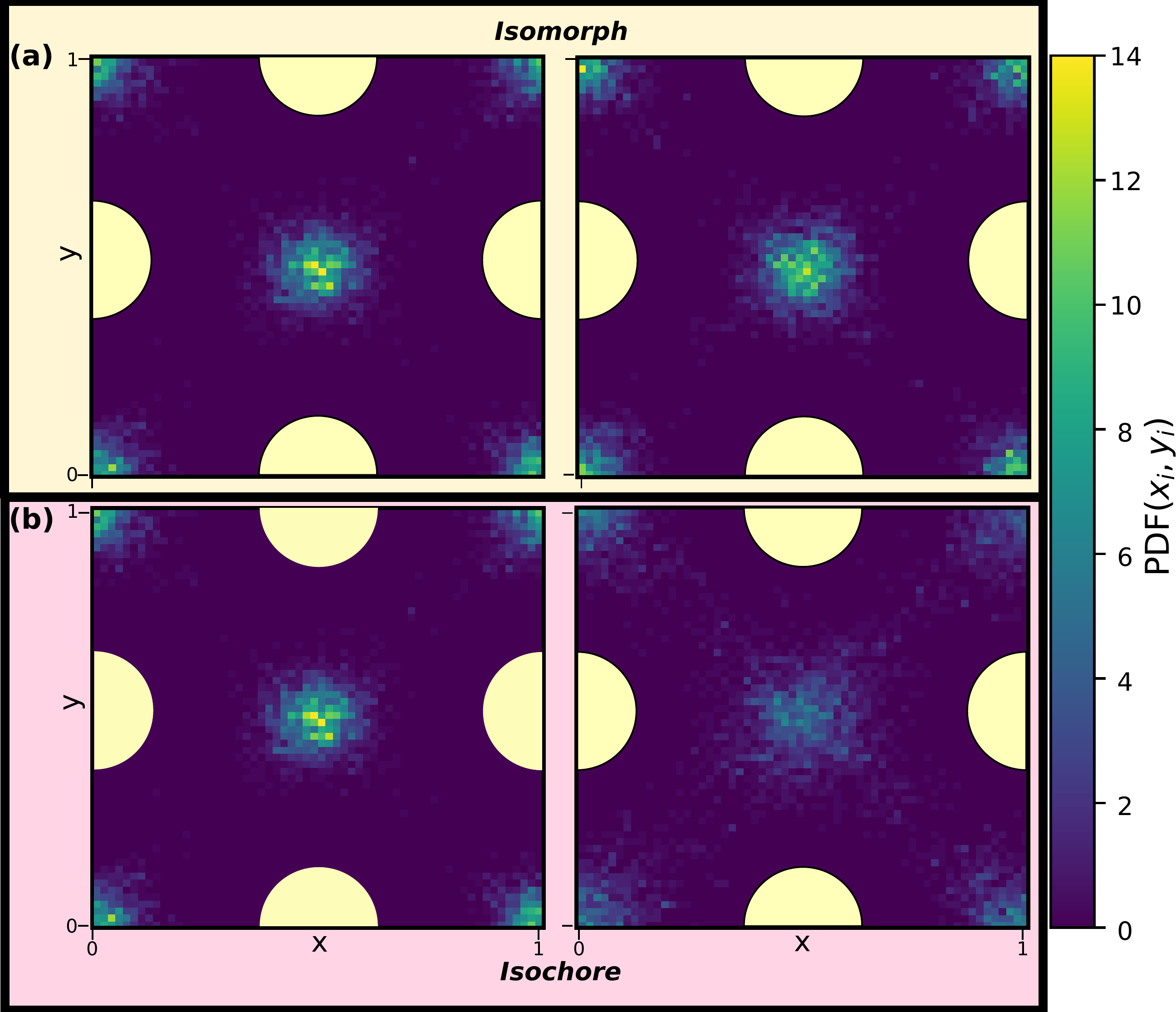}
  \caption{Reduced in-plane density profiles of the liquid particles in the layer closest to the walls, i.e. $|\widetilde{z}| \approx$ 4, for the first (left) and last (right) state points of the (a) isomorph (top) and (b) isochore (bottom) of Fig. \ref{fig5}. One
    unit cell of the crystalline walls is shown and the yellow circles indicate fcc wall particles.}\label{fig6}
\end{figure}

Figure \ref{fig6} displays the in-plane density profiles in the layer closest to the walls, i.e., for the layer around $|\widetilde{z}| \approx$ 4, for the first and the last state points of Fig. \ref{fig5}. The density profile is shown for one unit cell of the crystalline walls. The isomorph shows excellent scaling with the liquid particles situated in between the wall particles and exhibits very little change in the density profile whereas the isochore displays a density field that is increasingly smeared out as $T$ is increased, confirming the pre-crystallisation. Similar behavior is
observed the remaining layers (not shown).

\begin{figure}[H]
  \centering
  \includegraphics[width=140mm]{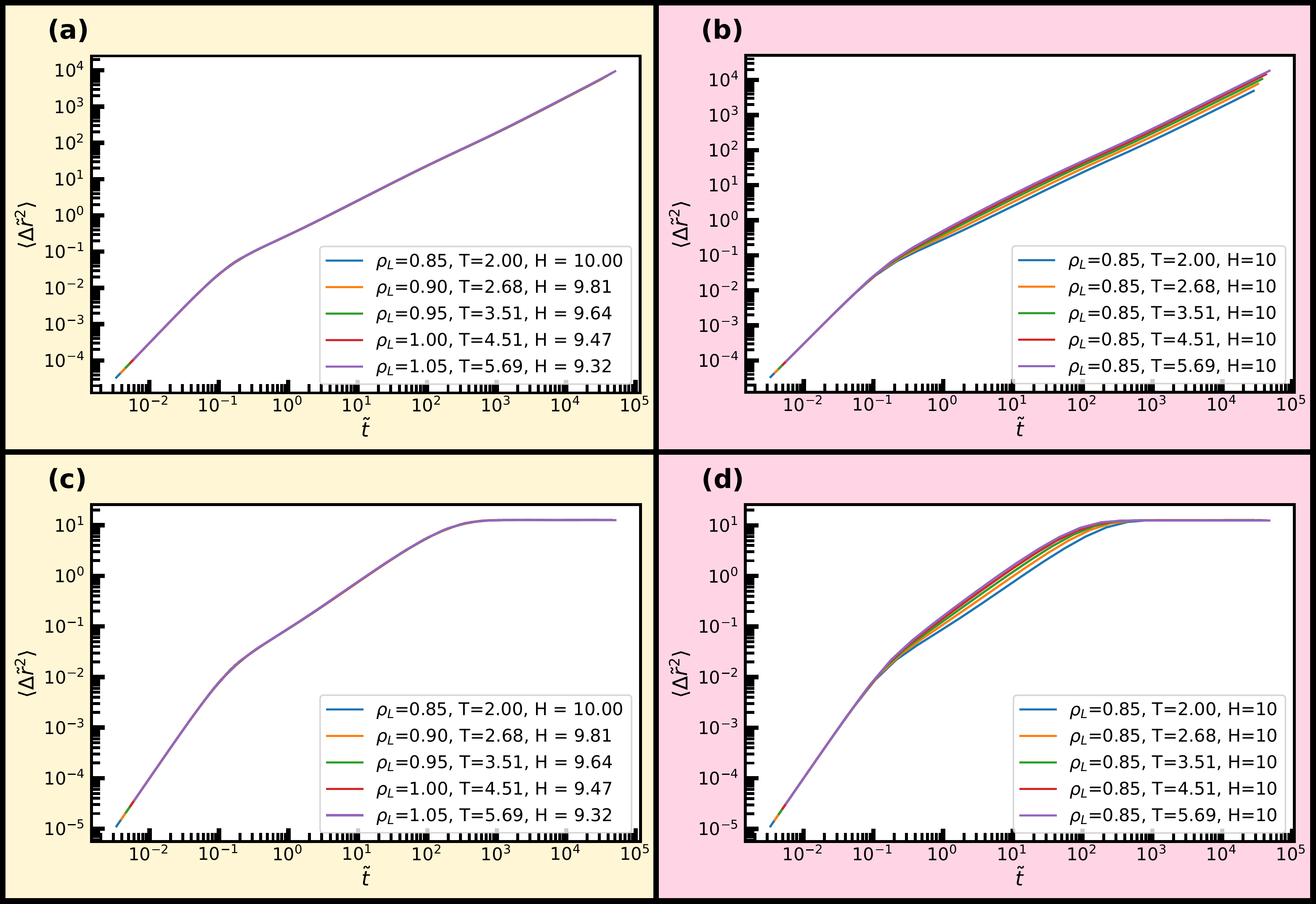}
  \caption{Reduced average mean-square displacements  parallel and normal to the walls along the isomorph and isochore of Fig. \ref{fig5}. (a) Isomorph, parallel dynamics. (b) Isochore, parallel dynamics. (c) Isomorph, normal dynamics. (d) Isochore, normal dynamics.}\label{fig7}
\end{figure}

Dynamical properties such as the mean-squared displacement (MSD) or the intermediate scattering function are also invariant along an isomorph and we now examine how well this behaviour holds in nanoconfinement. To do so,  we consider the reduced MSD parallel and normal to the walls as a function of reduced time in Fig. \ref{fig7} along the same isomorph and isochore as before. The
MSD is averaged over the entire slit-pore. We find excellent invariance along the isomorph and visible variation for the isochore. For the bulk liquid, at these high temperatures, one would see a
similar scaling in comparison to the isochore; the differences becoming more pronouced with the degree of supercooling.

\subsection{Isomorph invariance of higher-order structures}

Two-point spatial correlation functions, such as the radial distribution function, have been shown to a good approximation to be invariant along isomorphs in simulations \cite{paper4,ingebrigtsen2013,ingebrigtsen2015,exp1}. Higher-order structural correlations have been investigated to a lesser extent, and could be less invariant than the two-body correlation functions as the isomorph theory is approximate. In particular, geometric motifs, so-called \emph{locally favoured structures} (LFS) such as the bicapped square antiprism \cite{coslovich2007,malins2013fara} have been seen to vary by a factor of two along isomorphs in supercooled KABLJ \cite{malins2013iso}.

\begin{figure}[H]
  \centering
  \includegraphics[width=140mm]{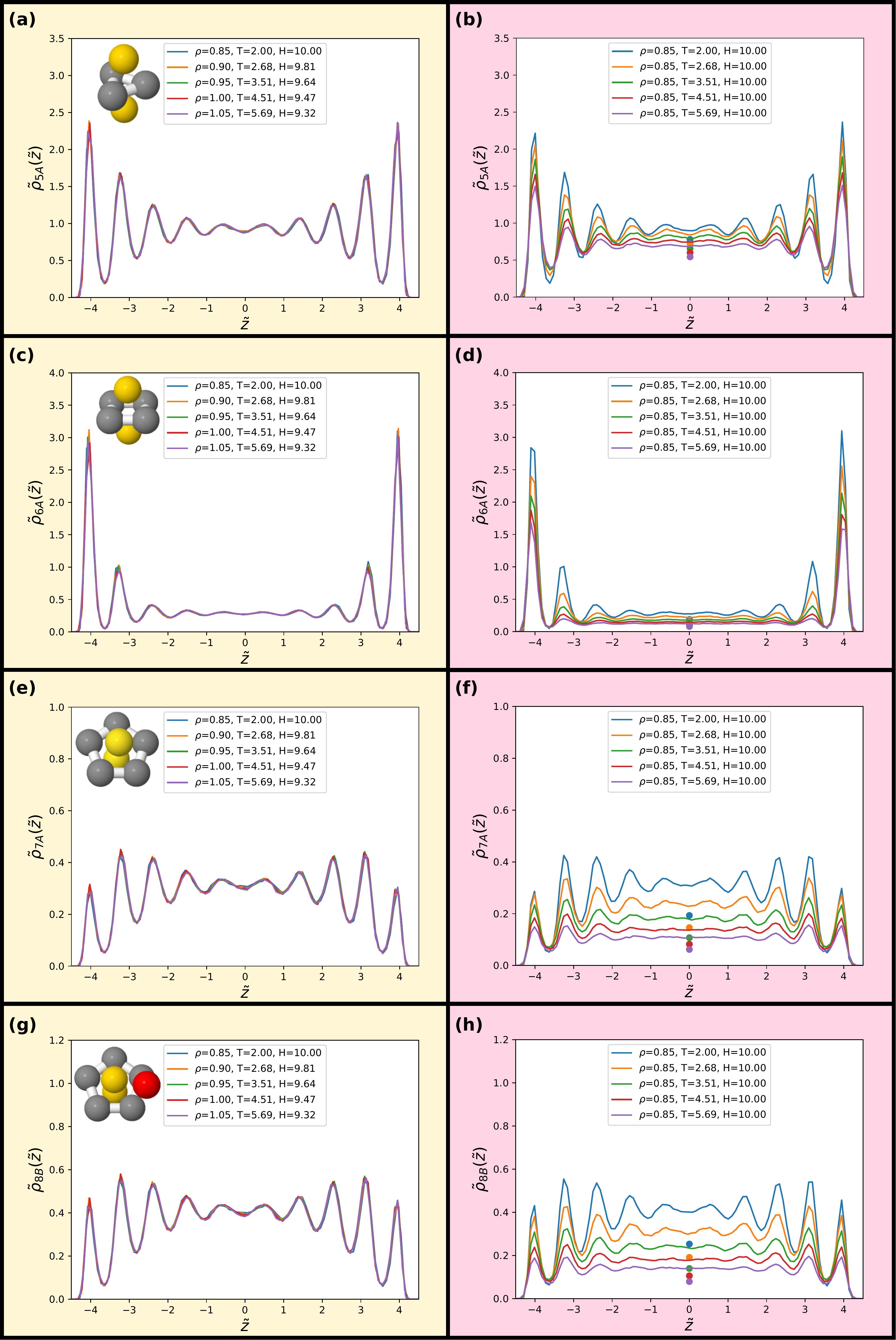}
  \caption{Populations of minimum energy clusters along the previously studied isomorph (left) and isochore (right). The minimum energy clusters
  considered in each case are illustrated in the corresponding panels. In particular, we consider 
  the 5-membered triangular bipyramid in (a) and (b), 
  the $m = 6$ octahedron in (c) and (d), 
  the $m = 7$ pentagonal bipyramid in (e) and (f) and 
  the $m = 8$ cluster with $C_s$ symmetry in (g) and (h). 
  The data points in (b,d,f,h)
  give bulk isochore values at the same state points.
  \label{fig8}}
\end{figure}

\begin{figure}[H]
  \centering
  \includegraphics[width=140mm]{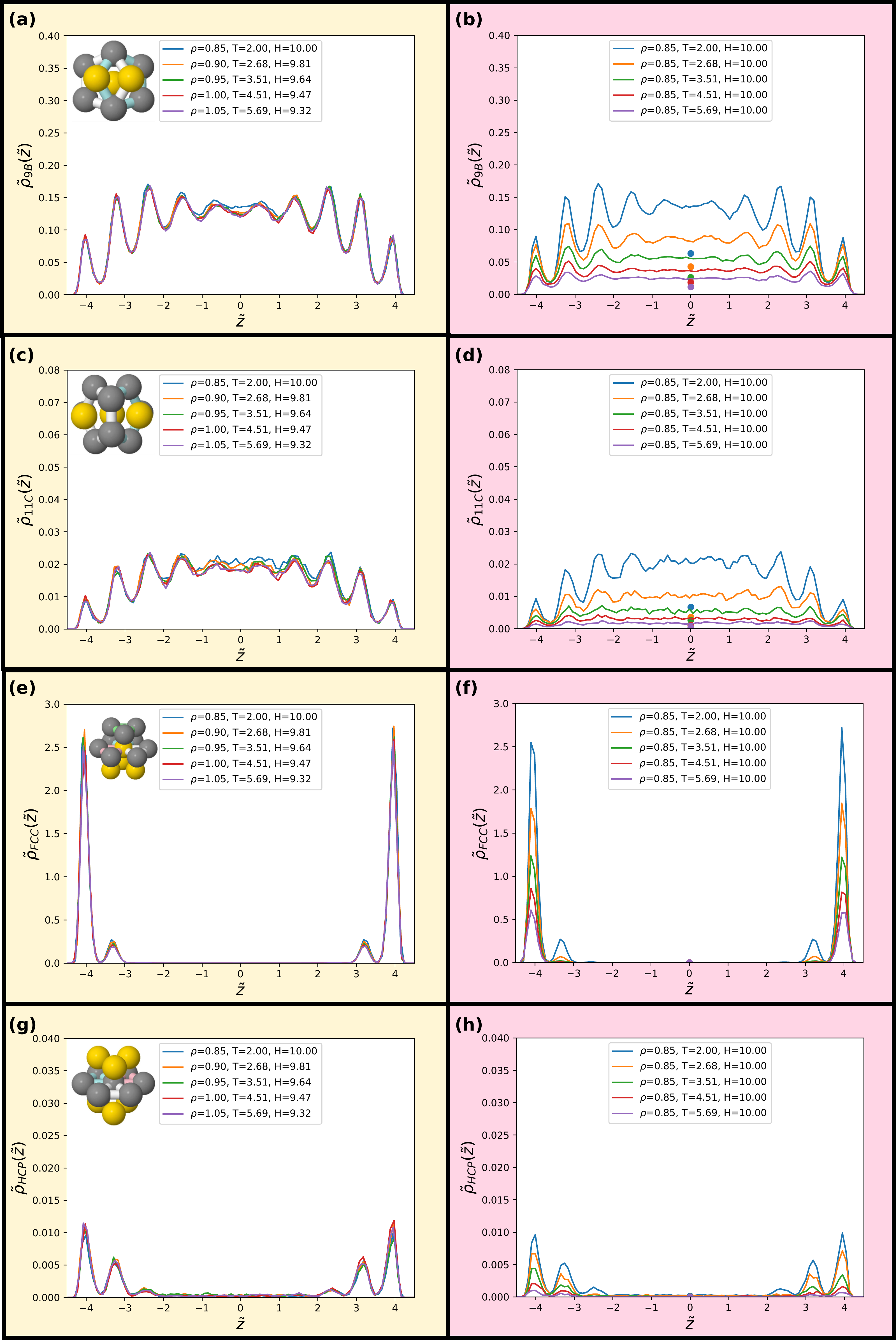}
  \caption{Populations of minimum energy clusters
  along the previously studied isomorph (left) and isochore (right). The minimum energy clusters
  considered in each case are illustrated in the corresponding panels. In particular, we consider 
  the 9-membered $C_{2v}$ symmetric cluster in (a) and (b), 
  the $m = 11$ $C_{2v}$ symmetric cluster in (c) and (d), 
  the fcc local crystalline environment in (e) and (f) and 
  the hcp local crystalline environment in (g) and (h).
  The data points in (b,d,f,h) 
  give bulk isochore values at the same state points.
\label{fig8a}}
\end{figure}

As a final probe of isomorphs for the confined SCLJ liquid we investigate in Figs. \ref{fig8} and \ref{fig8a} invariance of higher-order structures. Now the liquids here are not supercooled much and thus rather than a  locally favoured structure, we instead consider minimum energy clusters of $5\le m \le 13$ particles for these systems \cite{taffs2010,malins2013fara}. These minimum energy clusters typically include the locally favoured structure, although the latter typically exhibits a specific symmetry \cite{royall2015a} while the 
minimum energy clusters for each system exhibit a range of symmetries. In addition, we consider populations of the hcp and fcc crystalline structures.

The minimum energy structures are identified by the topological cluster classification \cite{malins2013tcc}. The topological cluster classification (TCC) algorithm carries out a Voronoi decomposition and seeks structures topologically identical to geometric motifs of particular interest. The
eight minimum energy structures of the SCLJ system
are depicted in the figures (top left in each panel) and which are minimum energy structures of the SCLJ system \cite{ben2018,malins2013tcc,taffs2010}.

We find that the distribution of clusters is, to an excellent approximation, invariant along the isomorph but not along the isochore. For all structures on the isochore the variation is around
a factor of two with hardly any visible variation along the isomorph. Very minor deviations are, however, noted for the 9B and 11C structures along the isomorph. We emphasize that the probing of these structures does not imply relevance of these structures for the dynamics of the liquid, but is merely used for testing invariance of higher-order structures along an isomorph.

While the structures considered exhibit very little change along the isomorph, the changes between the behaviour of the different clusters is notable in itself. We can identify three regimes: small amorphous clusters, larger amorphous clusters and crystalline structures. As shown in Figs. \ref{fig8}(a) and (c), the smaller amorphous clusters, the triangular bipyramid 5A and octahedron 6A largely follow the density profiles illustrated in Fig. \ref{fig5}. Larger amorphous clusters, beginning with the pentagonal bipyramid 7A have a degree of fivefold symmetry, and their population is suppressed close to the wall (see Figs. \ref{fig8}(e) and (g) and Figs. \ref{fig8a}(a) and (c)). Interestingly, this is \emph{not} observed in the case of a free interface where the cluster population is rather slaved to the density profile \cite{godonoga2011}. A rather different behaviour is found for the crystalline structures, where the layer by the wall has a high population of particles in a crystalline environment but the population in the middle of the slit is very small (Figs. \ref{fig8a}(e) and (g)). 

We compare these results with bulk populations of the same clusters for the same state points (temperature and density) as shown by the colored data points in Figs. \ref{fig8} and \ref{fig8a}. The data points are plotted for the isochore data, but may be taken to be representative of the isomorph data on the left hand side of the figure. Even in the centre of the slit, the results show a strong enhancement of cluster population with respect to the bulk in all cases except for the hcp and fcc crystals, whose population in the centre of the slit is negligible. This is remarkable, given that for a free liquid--vapour interface, cluster populations reach their bulk value with around a diameter from the interface \cite{godonoga2011}. Further work is called for to understand this unexpected increase in higher-order structure in confinement.

In the Supplementary Material (SM) we present results for an isomorph with $H \approx$ 6. We find also here
excellent scaling of density profiles and mean-square displacements along the isomorph. To conclude on the SCLJ liquid, we find that isomorphs survive into heavily confined systems with fcc crystalline walls and have excellent scaling even for higher-order structures.

\section{Kob-Andersen binary Lennard-Jones mixture}\label{res2}

We now turn to investigate similar quantities for the KABLJ mixture. Although being a binary mixture prized for its glassforming ability, the KABLJ mixture 
is prone to crystallization in the bulk \cite{ingebrigtsen2019}.
The mechanism for crystallisation in the bulk occurs through the formation of fcc (and hcp) nucleation of the majority A species \cite{ingebrigtsen2019}.
We find that the KABLJ mixture like the SCLJ liquid in confinement also crystallizes in the layer closest to the wall
and suggests that heterogeneous nucleation at the walls (which are patterned as an fcc structure) is indeed a powerful mechanism also for the KABLJ mixture as seen for other
simple liquids \cite{tox1981}. 

\subsection{Variation in the correlation coefficient $R$ and density-scaling exponent $\gamma$}

\begin{figure}[H]
  \centering
  \includegraphics[width=140mm]{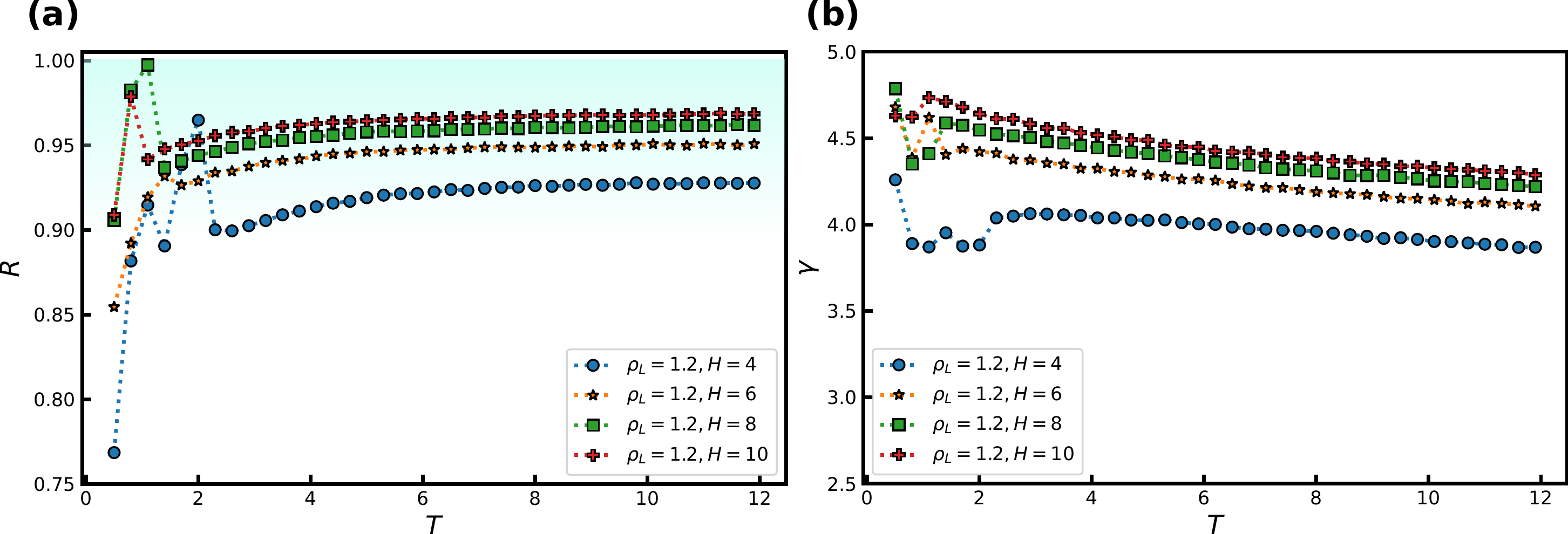}
  \caption{$R$ and $\gamma$ as a function of temperature $T$ for several slit-pore widths $H$ for the KABLJ mixture at $\rho_{\rm L}$ = 1.2 and $\rho_{\rm W}$ = 1. (a) $R$ as a
      function of temperature. (b) $\gamma$ as a function of temperature.}\label{fig13}
\end{figure}

Figure \ref{fig13} displays $R$ and $\gamma$ as a function of temperature and several slit-pore widths $H$ for the KABLJ mixture at $\rho_{\rm L}$ = 1.2 and $\rho_{\rm W}$ = 1 with $\epsilon_{\rm A W}$ = 1,
and $\epsilon_{\rm B W}$ = 0.707. The bulk correlation coefficient at this density is $R \approx 0.96$. As for the SCLJ liquid we find an increase in $R$ and a decrease in $\gamma$ with temperature. Similary $R$ increases with increasing $H$ but even for $H = 4 $ is the confined liquid R-simple. 

\begin{figure}[H]
  \centering
  \includegraphics[width=140mm]{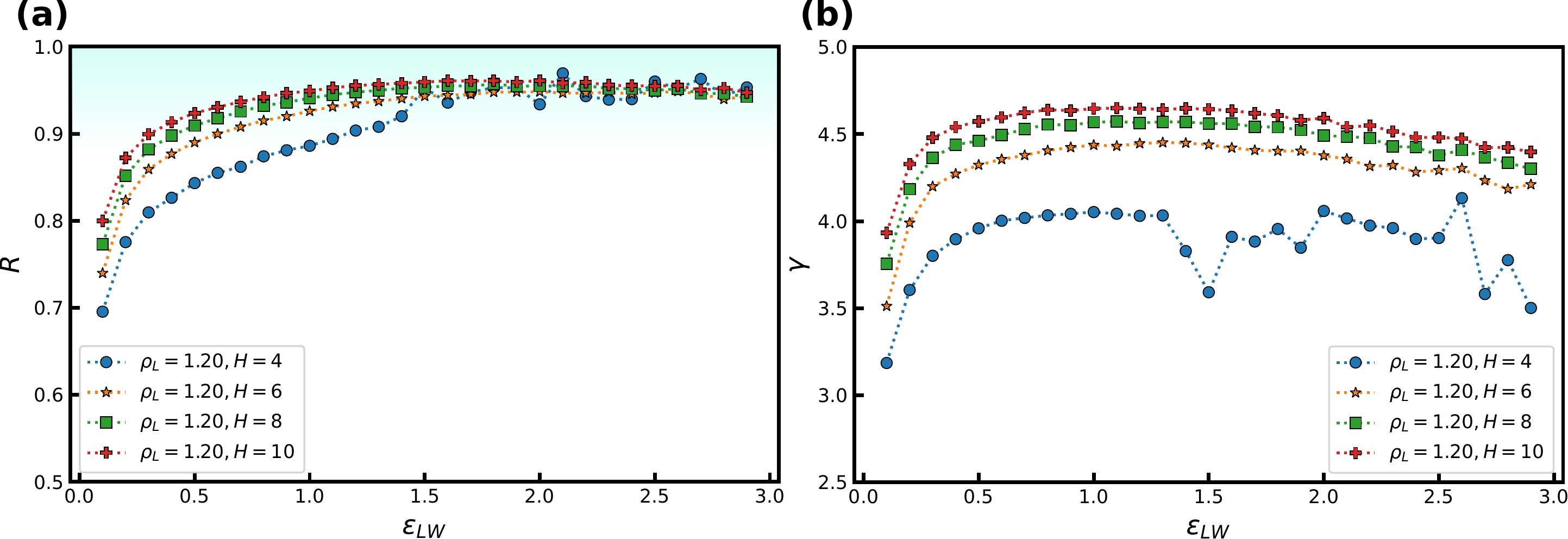}
  \caption{$R$ and  $\gamma$  as a function of the liquid-wall interaction strength $\epsilon_{\rm LW}$ for several slit-pore widths $H$. Temperature is fixed at $T$ = 2.5 and the
    liquid density is $\rho_{\rm L}$ = 1.2. (a) R as a function of $\epsilon_{\rm LW}$. (b) $\gamma$ as a function of $\epsilon_{\rm LW}$.}\label{fig14}
\end{figure}

Next, we consider the effect of changing the liquid-wall interaction strength $\epsilon_{\rm LW}$ in Fig. \ref{fig14}. A maximum is again noted for both $R$ and $\gamma$ and is displaced away
from the value of $\epsilon_{\rm A W} = 1$; The location of the maximum seems to be more dependent on $H$ than for the SCLJ.
For the KABLJ mixture the effect of surface roughness $\rho_{\rm W}$ was not investigated.

\subsection{Isomorph invariance of reduced density profiles and dynamics}

\begin{figure}[H]
  \centering
  \includegraphics[width=140mm]{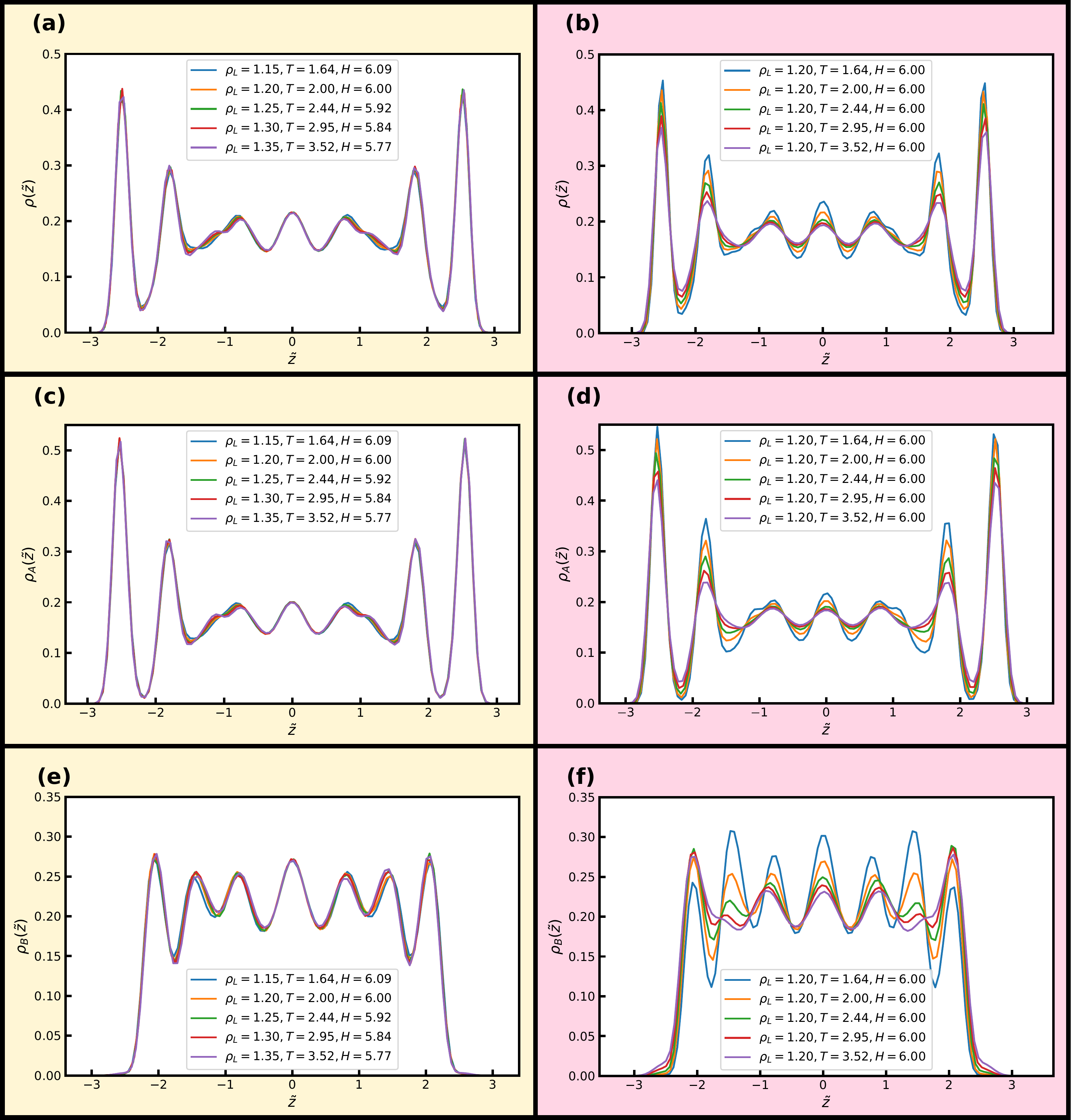}
   \caption{Reduced density profiles of the liquid particles perpendicular to the walls along an isomorph and an isochore. (a) Total density profile, isomorph. (b) Total density profile, isochore. (c) A-particle density profile, isomorph. (d) A-particle density profile, isochore. (e) B-particle density profile, isomorph. (f) B-particle density profile, isochore.}\label{fig15}
\end{figure}

For the KABLJ mixture we also investigate two isomorphs with $H \approx$ 6 and $H \approx 10$ to facilitate comparison with the SCLJ liquid in the previous section.
Figures \ref{fig15}(a), (c), and (e) show reduced normal density profiles for both particles in the mixture ($A$ and $B$), as well as the total density along an isomorph with 17$\%$ density increase and $H \approx 6$. Figures \ref{fig15}(b), (d), and (f) show the corresponding quantities along
an isochore. Good invariance is noted along the isomorph but not along the isochore where again significant changes in the peak heights are noted for all density profiles, in particular for the $B$-particle
density profiles (Fig. \ref{fig15}(f)).

\begin{figure}[H]
  \centering
  \includegraphics[width=120mm]{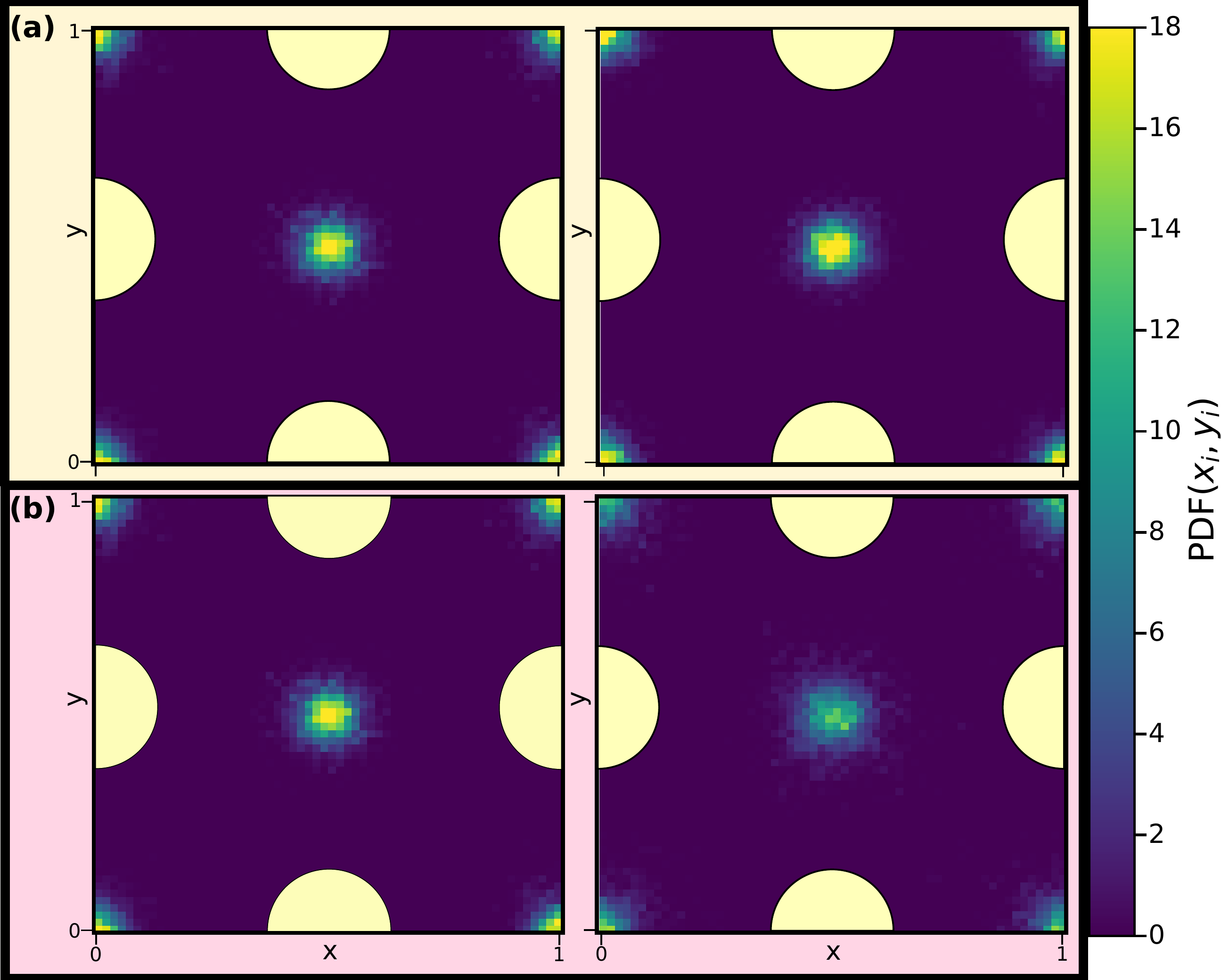}
  \caption{Reduced in-plane total density profiles of the liquid particles in the layer closest to the walls, i.e. $|\widetilde{z}| \approx$ 2.7, for the first (left) and last (right) state points along the isomorph (top) and the isochore (bottom) of Fig. \ref{fig15}. One
    unit cell of the crystalline walls is shown with the yellow circles indicating fcc wall particles.}\label{fig16}
\end{figure}

Figure \ref{fig16} shows reduced in-plane total density profiles for the layer closest to the wall, i.e. $|\widetilde{z}| \approx$ 2.7, for the first and last state points of the same isomorph and isochore.
For the KABLJ mixture the in-plane density profile does not seem to show a strong deterioration along the isochore as found for the SCLJ liquid, which could be
anticipated from the height variation of the first peak in Fig. \ref{fig15}(b). Nevertheless the invariance is still visually worse than the isomorph.

\begin{figure}[H]
  \centering
  \includegraphics[width=140mm]{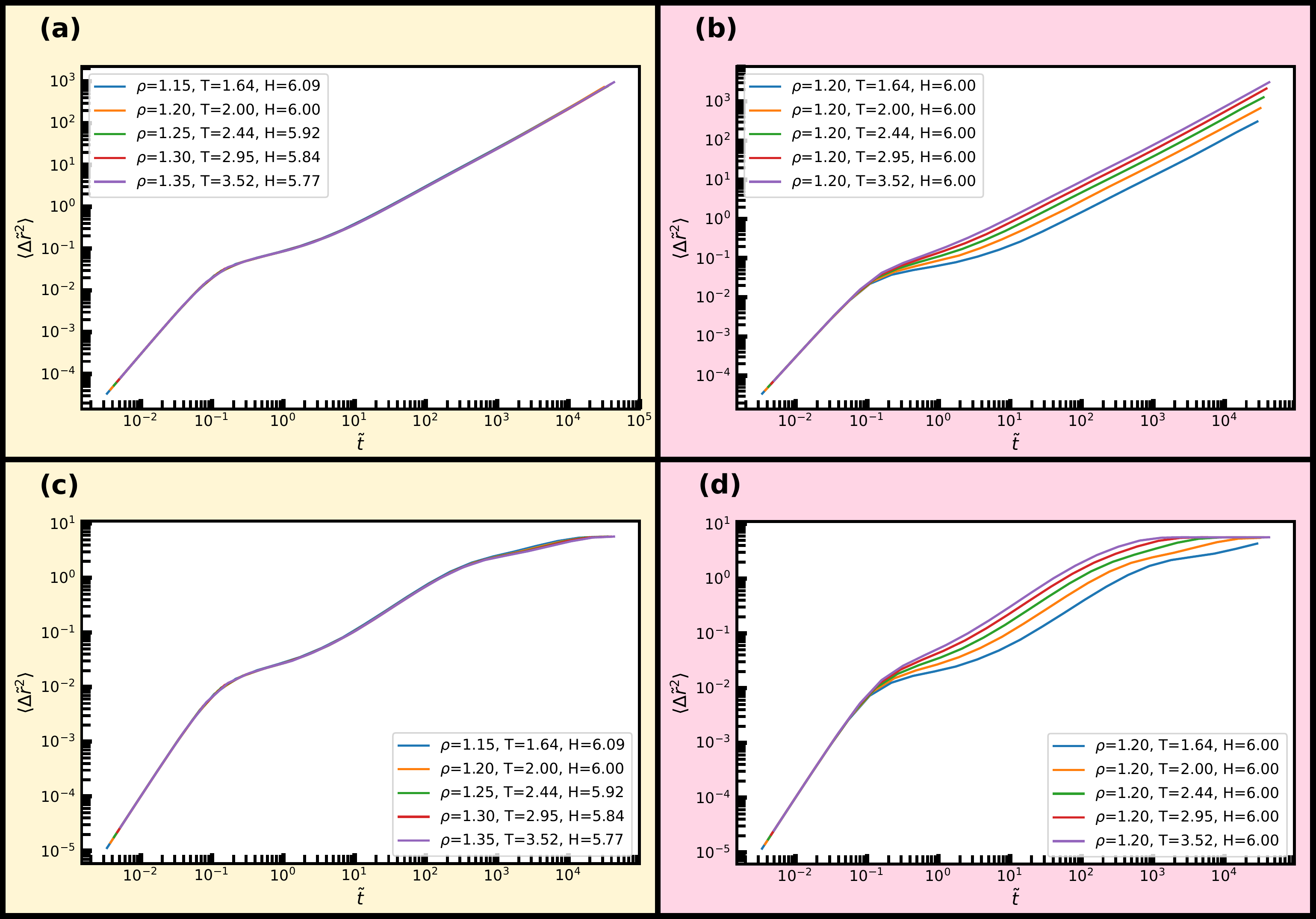}
  \caption{Reduced A-particle mean-square displacement parallel and normal to the walls averaged over the entire slit-pore along an isomorph and an isochore. (a) Isomorph, parallel dynamics. (b) Isochore, parallel dynamics. (c) Isomorph, normal dynamics. (d) Isochore, normal dynamics.}\label{fig17}
\end{figure}

For the reduced normal and parallel A-particle MSDs in Fig. \ref{fig17} almost perfect scaling is observed along the isomorph while approximately a decade deviation in diffusion coefficient is observed for the isochore. These deviations in MSD are similar to what is seen for supercooled bulk liquids \cite{paper4,ingebrigtsen2015}. We find for the B-particles a very similar scaling behaviour (not shown). 

\subsection{Isomorph invariance of higher-order structures}

\begin{figure}[H]
  \centering
  \includegraphics[width=140mm]{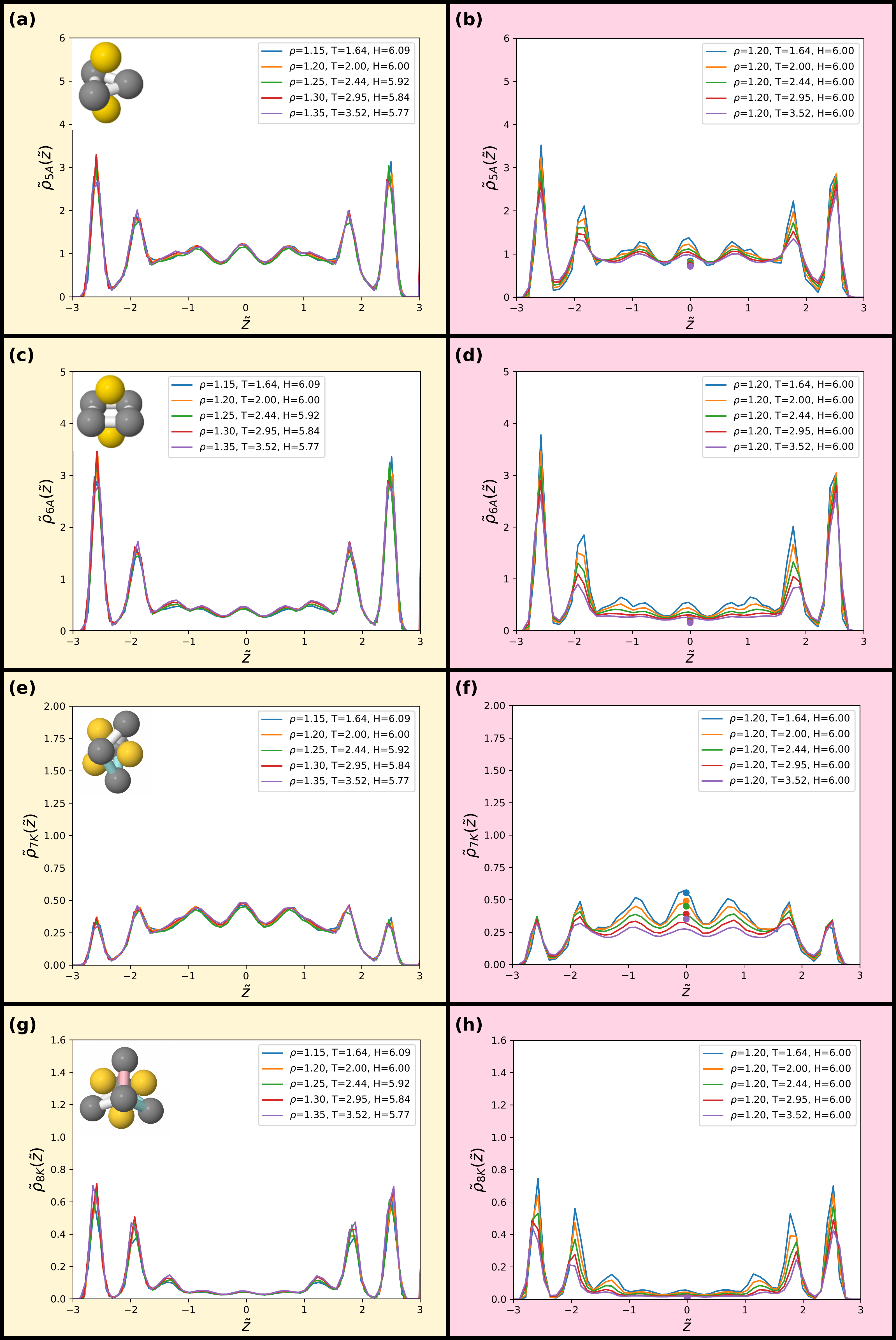}
 \caption{Populations of minimum energy clusters
  along the previously studied isomorph (left) and isochore (right). The minimum energy clusters
  considered in each case are illustrated in the corresponding panels. In particular, we consider 
  the 5-membered triangular bipyramid in (a) and (b), 
  the $m = 6$ octahedron in (c) and (d), 
  the $m = 7$ polytetrahedron 
  in (e) and (f) and 
  the $m = 8$ pyramidal geometry 
  in (g) and (h).
  The data points in (b,d,f,h)
  give bulk isochore values at the same state points.
  \label{fig18}}
\end{figure}

\begin{figure}[H]
  \centering
  \includegraphics[width=140mm]{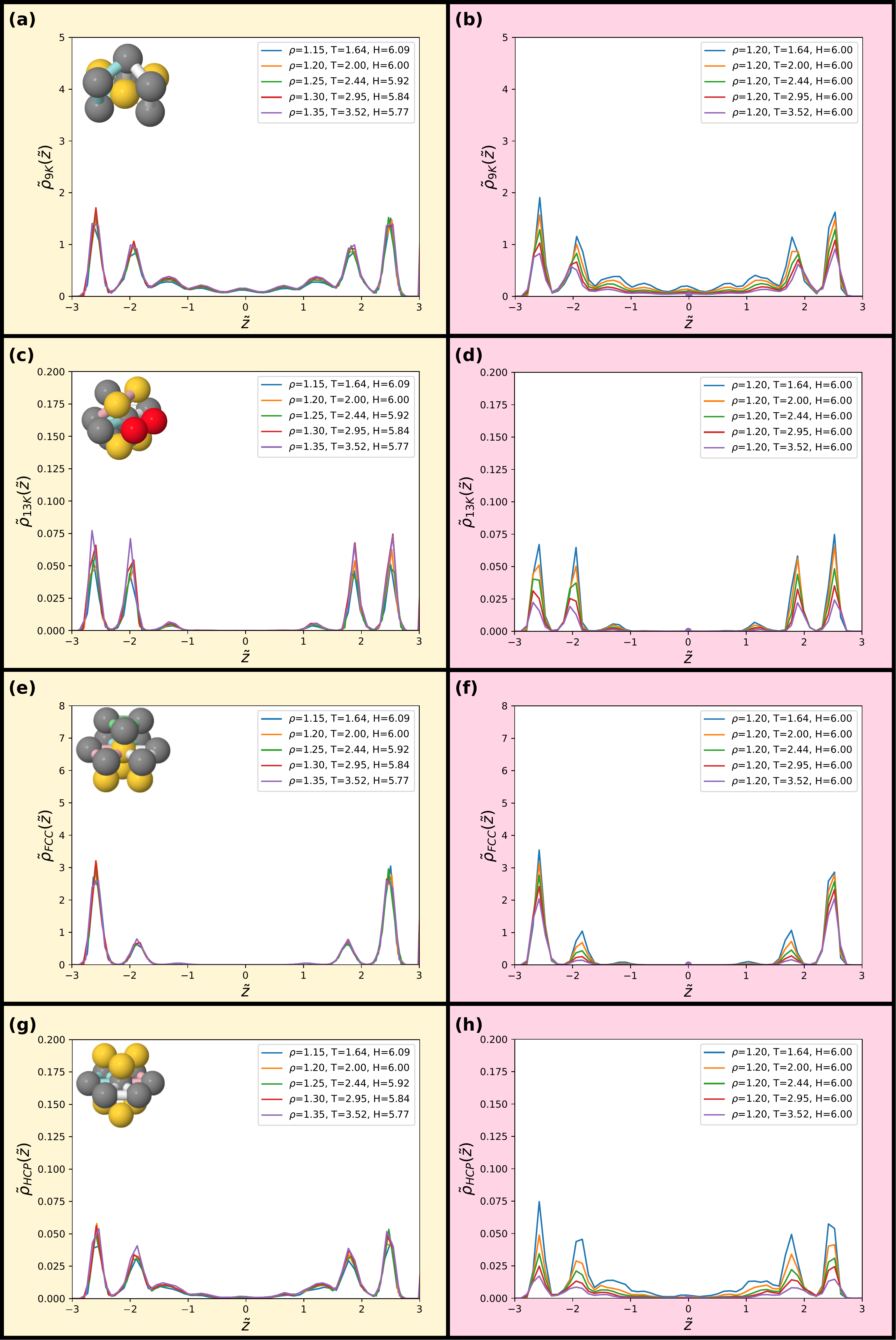}
  \caption{Populations of minimum energy clusters
  along the previously studied isomorph (left) and isochore (right). The minimum energy clusters
  considered in each case are illustrated in the corresponding panels. In particular, we consider 
  the 9-membered triangular antiprism in (a) and (b), 
  the $m = 13$  polytetra-octahedron (c) and (d), 
  the fcc local crystalline environment in (e) and (f) and 
  the hcp local crystalline environment in (g) and (h). 
  The data points in (b,d,f,h)
  give bulk isochore values at the same state points.
  }\label{fig18a}
\end{figure}

Finally, we consider invariance of selected minimum energy clusters
for the KABLJ mixture \cite{malins2013tcc,malins2013fara} in Figs. \ref{fig18} and  \ref{fig18a}, reaching a similar conclusion as for the SCLJ liquid
with excellent invariance along the isomorph but not along the isochore showing around a factor of two variation in almost all structures. Although, the bicapped square antiprism (11A) structure has been shown to correlate reasonably to the dynamics of the KABLJ system \cite{peter2015,malins2013fara,speck2012,turci2017} we find very few bicapped square antiprisms
at the higher temperatures considered here (the onset temperature for glassy dynamics, at which the bicapped square antiprism become popular is around $T_\mathrm{on}\approx1.0$) and it is therefore not included in the figures. 

In particular, it is quite remarkable that the minimum energy clusters of the KABLJ system \cite{malins2013fara,malins2013tcc} show such a good invariance, even for higher-order correlations, given that previous work showed a  significant discrepancy in precisely the same system \cite{malins2013iso}, although at a lower temperature $T$ = 1.0. The results presented here is at a higher tempererature ($T >$ 2.0), the magnitude of the discrepancy, and its rather weak temperature dependence leads one to speculate whether the confinement may somehow influence the agreement. 

In comparing with the bulk values (data points in Figs. \ref{fig18} and  \ref{fig18a} right hand side), we see that as above in the case of the SCLJ system (Figs. \ref{fig8} and  \ref{fig8a}) that in many cases the cluster population even at the centre of the slit is markedly higher than the bulk liquid at the same state point. However this is not universally the case here, as the $m=7$ polytetrahedron (7K) in fact seems to sit right on the confined data for some state points, while at higher temperature the bulk population seems rather higher than the confined system of interest here. The reasons for the change in behaviour of this structure and, as noted above, why the minimum energy clusters typically have a reduced population with respect to the bulk is an interesting topic for future work.

We provide in SM figures for an isomorph with $H \approx$ 10. Similar conclusions are reached showing again good invariance along the isomorph.

\section{Summary and outlook}\label{con}

Isomorphs are curves in the thermodynamic phase diagram of R-simple liquids along which 
structure and dynamics in reduced units to a good approximation are invariant. However, nanoconfined liquids show strikingly different behavior from bulk liquids in terms of their structure and dynamics \cite{schoen1987}. It is therefore not obvious that concepts demonstrated in the bulk apply also to confined liquids. Extending the isomorph framework to nanoconfined liquids is important as theories for nanoconfined fluids have been slower develop due to the added complexity. 

A previous study \cite{ingebrigtsen2013} explored the existence of isomorphs in nanoconfined liquids using smooth slit-pore geometry
and found that isomorphs
do survive under confinement.
Here, we have studied the effect of introducing highly inhomogenous density profiles both parallel and perpendicular to the walls by applying more realistic crystalline fcc walls. The effect of the wall-to-wall distance, the strength of liquid-wall interactions $\epsilon_{\rm LW}$, and surface roughness $\rho_{\rm W}$ were explored on two bulk R-simple liquids: the SCLJ liquid and the KABLJ mixture.

Although strong inhomogeneities occur in this type of confinement, we found that R-simple liquids and isomorphs survive down to a few particle diameters confinement. More specifically, we probed
density profiles, normal and parallel mean-squared displacements, as well as higher-order structures using the topological cluster classification algorithm along isomorphs. Even for higher-order correlations of populations of minimum energy clusters up to 13 particles, we find excellent invariance along the isomorphs. This is notable, as in the bulk, albeit at lower temperatures, considerable deviation was found for higher-order structures even when the two-point structure appeared to scale well \cite{malins2013iso}. Curiously, in many (but not all) of these clusters, the population in the confined system, even at the centre, is markedly higher than that in the bulk, even for the same state point. This is remarkable, given that in the case of a free liquid-vapour interface, the cluster population reaches its bulk value within around a diameter of the interface \cite{godonoga2011}. This curiosity will be investigated in future work.

We conjecture from the current study that even very complicated confinements, e.g., carbon nanotubes, can exhibit Roskilde simplicity and thus also their associated scaling laws, such as Rosenfeld excess-entropy scaling. This provides an important simplification of the phase diagram and valuable insights into the structures and dynamics of confined liquids. In this connection, an intriguing path for further research is the development of equations of state for confined liquids using the isomorph theory \cite{paper5}.

\subsection*{Conflicts of interest}
There are no conflicts of interest to declare

\subsection*{Acknowledgements}

B.M.G.D.C., J.C.D., and T.S.I. are supported by the VILLUM Foundation’s Matter Grant (No. 16515). B.M.G.D.C. acknowledges funding from EPSRC with grant code EP/L016648/1. C.P.R. acknowledges the Royal Society and European Research Council (ERC consolidator grant NANOPRS, project number 617266) for financial support. 


\end{document}